\documentclass[aps,prl,preprint,groupedaddress,showpacs]{revtex4-1}
\usepackage{graphicx}
\usepackage{color}
\usepackage{amssymb,amsfonts,amsmath,mathrsfs}

\begin{document}

%\preprint{}

\title{Robustness of spatial patterns in buffered reaction-diffusion systems\\ and its reciprocity with phase plasticity}


\author{Tetsuhiro S. Hatakeyama}
\email[]{hatakeyama@complex.c.u-tokyo.ac.jp}
\author{Kunihiko Kaneko}
%\thanks{}
\affiliation{Department of Basic Science, University of Tokyo, \\
3-8-1 Komaba, Meguro-ku, Tokyo 153-8902, Japan}


\date{\today}

\begin{abstract}
Robustness of spatial pattern against perturbations is an indispensable property of developmental processes for organisms, which need to adapt to changing environments. Although specific mechanisms for this robustness have been extensively investigated, little is known about a general mechanism for achieving robustness in reaction-diffusion systems. Here, we propose a buffered reaction-diffusion system, in which active states of chemicals mediated by buffer molecules contribute to reactions, and demonstrate that robustness of the pattern wavelength is achieved by the dynamics of the buffer molecule. This robustness is analytically explained as a result of the scaling properties of the buffered system, which also lead to a reciprocal relationship between the wavelength's robustness and the plasticity of the spatial phase upon external perturbations. Finally, we explore the relevance of this reciprocity to biological systems.
\end{abstract}

%\keywords{}

\maketitle


Robustness is ubiquitous in biological systems. Developmental processes are robust to environmental changes, a property described as canalization \citep{Waddington1942}. The robustness of each individual developmental process, including regulation in signaling, cellular differentiation, and pattern formation, has been analyzed both experimentally and theoretically. In particular, the robustness of pattern formation by reaction-diffusion (RD) dynamics has been studied in relationship to proportion preservation: Although the body size of organisms varies, the proportion of the size of each organ to the whole body is conserved, as is the robustness of the number of body segments. Ordinary pattern formation systems, however, have their own characteristic wavelengths \citep{Turing1952, Kuramoto1984}, and do not preserve proportions per se. Thus, the mechanisms of proportion regulation have been investigated, both theoretically and experimentally \citep{Othmer1980, Hunding1988, Barkai2009, Inomata2013, Werner2015}.

Spatial pattern robustness is also important for cellular polarity. It is the result of a cellular compass \citep{Weiner2002}, which is believed to be generated by an RD process on the cell membrane \citep{Mori2008}. Then, the number of the cellular compass, given as a pattern wavelength in the RD system, must be maintained against environmental changes for robust cellular polarity.
 
Another example of robust pattern formation is seen in the configuration of differentiated cells. In multicellular cyanobacteria, e.g., {\it Anabaena} and {\it Nostoc}, cells are linked like beads on a string, where some differentiate from vegetative cells to heterocysts under nitrogen-depleted condition \citep{Kumar2010}. Vegetative cells can fix carbon from carbon dioxide by photosynthesis, while heterocysts can only fix nitrogen in the atmosphere, as nitrogen fixation and photosynthesis are biochemically incompatible \citep{Agapakis2012}. When the nitrogen level in culture media decreases, approximately one tenth of cells differentiate into heterocysts, which form a spatially periodic pattern. Although intracellular processes crucially depend on nitrogen concentration, the frequency of heterocysts is preserved against its change, as confirmed for some species of {\it Nostoc} \citep{Vintila2007}.

In general, the characteristic length of patterns generated by RD systems depends on the ratio between the characteristic timescale of the reaction and diffusion coefficients. Reaction speeds and diffusion coefficients generally have different dependences on environmental conditions (e.g., temperature or external chemical concentration). Hence, the characteristic length of a pattern will change unless some robustness mechanism exists to preserve it. Indeed, in proportion preservation, reaction speeds are expected to be regulated to counterbalance body-size changes \citep{Othmer1980, Hunding1988, Barkai2009, Inomata2013, Werner2015}. In this letter, we propose a general mechanism for pattern robustness, where some buffer molecules regulate reaction speed and counterbalance environmentally induced changes.

A related, but distinct property that is essential to biological systems is plasticity. Although robustness and plasticity might seem incompatible, we previously showed that period robustness and phase plasticity are reciprocal in biological clocks \citep{Hatakeyama2015}.
Here, using the above buffering mechanism, we generalize this reciprocity.
Specifically, we show that systems with more robust spatial pattern wavelengths have higher phase plasticity in spatial patterns, i.e., they are more changeable by transient environmental perturbations.
We then provide a unified description of the reciprocities in spatial pattern and temporal rhythm, and discuss the relevance of this reciprocity to biological systems.

We introduce a model that can show spatial pattern formation with RD dynamics with buffer molecules, which can counterbalance environmental changes by altering their concentrations.
Here we consider the following RD system:
\begin{subequations}
\begin{eqnarray}
\frac{\partial x_i}{\partial t} &=& f_i(\{x_j\}, \{w_j\}; \beta) + D_{x_i} \frac{\partial^2}{\partial {\bf r}^2} x_i, \\
\frac{\partial w_i}{\partial t} &=& g_i(\{x_j\}, \{w_j\}; \beta) + D_{w_i} \frac{\partial^2}{\partial {\bf r}^2} w_i,
\end{eqnarray}
\end{subequations}
where $x_i$ is the concentration of the $i$th molecule species that forms a spatial pattern by RD dynamics and $w_i$ is the concentration of the $i$th buffer molecule.
$D_{x_i}$ and $D_{w_i}$ are diffusion constants of $x_i$ and $w_i$, respectively, $\beta$ is an environmental factor, (e.g., temperature or the concentration of a nutrient molecule). 
Here, we assume that components for pattern formation, $x_i$, and for the buffer, $w_i$, are separate, for the sake of simplicity.
However, even if they are not, an approximate form of Eq.(1) can be adopted by suitable variable transformation.

\begin{figure}[htbp]
\centering \includegraphics{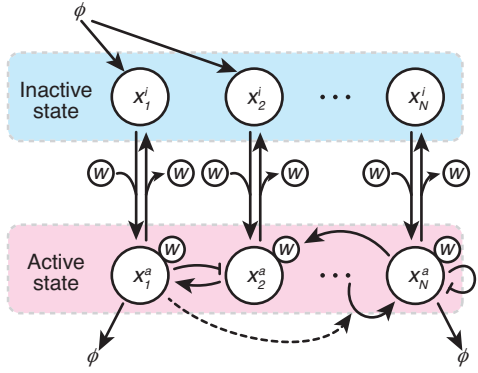}
\caption{
Schematic of a buffered reaction-diffusion system. $x_i^i$ and $x_i^a$ are inactive and active $i$th molecules, respectively. $x_i^i$ turns into $x_i^a$ by binding with a buffer molecule $w$.
Reactions occur only among active molecules.
Solid, flat-headed, and dotted arrows represent activation, inhibition, and catalysis, respectively.}
\label{scheme}
\end{figure}

Here, we consider a system in which one buffer molecule is sufficient for the robustness of the pattern wavelength.
As shown in Fig.\ref{scheme}, molecules $x_i$ have two states, active and inactive, denoted by $x_i^a$ and $x_i^i$, and the state change is mediated by binding with the buffer molecule, as $x_i^i + w \rightleftharpoons x_i^a$
(see also\citep{Hunding1988, Lengyel1992} for the use of such additional molecule for RD system).
Next, reactions of $x_i$ for pattern formation are assumed to take place only between active molecules.
We also assume that dependence on the environmental parameter is given by the environment-sensitivity function (ESF), $\mathscr{F}(\beta)$, which is assumed to be separable from the reaction functions.
Here, dynamics of buffer is given by $g(w; \beta)$ independent of $x_i$s.
Then, the dynamics are given by
\begin{subequations}
\begin{eqnarray}
\frac{\partial x_i}{\partial t} &=& \mathscr{F}(\beta) f_i(\{x_{j}^a\}) + D_{x_i} \frac{\partial^2}{\partial {\bf r}^2} x_i, \label{dxdt} \\
\frac{\partial w}{\partial t} &=& g(w; \beta) + D_{w} \frac{\partial^2}{\partial {\bf r}^2} w.
\end{eqnarray}
\end{subequations}
Assuming that the association and dissociation reactions between the buffer and other molecules are faster than other reactions required for pattern formation, $w$ relaxes to the steady state given as $w^*(\beta)$ at $\beta$, where $x_i^a$ is adiabatically eliminated as $x_i^a = x_i w / (w + K_{x_i})$,  as in Michaelis-Menten kinetics under the conservation condition $x_i$, as $x_i^a + x_i^i = x_i$.

One of the simplest examples of $g(w; \beta)$ that naturally meets the above criteria is $g(w; \beta) = \mathscr{G}_1 (\beta) - \mathscr{G}_2 (\beta) w$, where $\mathscr{G}_1 (\beta)$ and $\mathscr{G}_2 (\beta)$ are ESFs, and $w^*(\beta)$ is given as $w^*(\beta) = \mathscr{G}_1 (\beta) / \mathscr{G}_2 (\beta)$.
As an ESF, the Arrhenius equation is an example where $\beta$ is the inverse temperature with a unit of the Boltzmann constant as unity, i.e., $\mathscr{F}(\beta) = \exp(-\beta E_f)$, with $E_f$ as an activation energy.

We now derive a scaling property of the buffered RD system to show that wavelength robustness only holds when buffer concentration $w^*$ is less than any dissociation constants $K_{x_i}$. 

First, when the steady-state concentration of $w$ is higher than any $K_{x_i}$s, $x_i^a$ can be approximated as $x_i^a \sim x_i$. Hence, Eq.(2) can be rescaled by transformation of spatial scale as ${\bf r} \rightarrow {\boldsymbol \rho} / \mathscr{F} (\beta)^{1/2}$ and timescale as $t \rightarrow \tau / \mathscr{F} (\beta)$, given as
\begin{equation}
\frac{\partial x_i}{\partial \tau} = f_i(\{x_j\}) + D_{x_i} \frac{\partial^2}{\partial {\boldsymbol \rho}^2} x_i.
\end{equation}
Due to the transformation of spatial scale, the wavelength is proportional to $1 / \mathscr{F} (\beta)^{1/2}$ and is not robust against change in $\beta$, as in the case without buffer molecules.

In contrast, when the steady-state concentration of $w$ is lower than any $K_{x_i}$s, $x_i^a$ can be approximated as $x_i^a \sim w x_i / K_{x_i}$. In this case, Eq.(2) can be rescaled by transformation of spatial scale as ${\bf r} \rightarrow {\boldsymbol \rho} / \left\{ \mathscr{F} (\beta) w^* (\beta) \right\}^{1/2}$, with the timescale change $t \rightarrow \tau / \mathscr{F} (\beta) w^* (\beta)$, and concentration $x_i \rightarrow \tilde{x}_i / w^* (\beta)$:
\begin{equation}
\frac{\partial \tilde{x}_i}{\partial \tau} = F_i(\{\tilde{x}_j\}) + D_{x_i} \frac{\partial^2}{\partial {\boldsymbol \rho}^2} \tilde{x}_i, \label{scaledBuffer}
\end{equation}
where $F_i(\{ \tilde{x}_j \}) = f_i(\{ \tilde{x}_j / K_{x_i} \})$ \citep{foot1}.
Due to the transformation in spatial scale, the wavelength is proportional to $1 / \left\{ \mathscr{F} (\beta) w^* (\beta) \right\}^{1/2}$ and the relative change in wavelength is given as $\Delta \ln \lambda = - \left\{ \Delta \ln \mathscr{F} (\beta) + \Delta \ln w^* (\beta) \right\} / 2$. Thus, the wavelength change can be compensated for by the steady-state concentration of the buffer molecule if $\Delta \ln \mathscr{F}(\beta) + \Delta \ln w^* (\beta) = 0$.
Hence, if the steady-state concentration of $w$ is regulated as $w^*(\beta) \propto 1 / \mathscr{F}(\beta)$, the wavenumber is robust to $\beta$.
There could be several ways to achieve it.
For example, if the buffer flows in with a constant speed and is degraded following $\mathscr{F}(\beta)$, the above criterion is satisfied.
Even if such condition is not satisfied, robust patterning can be achieved by balancing the synthesis and degradation of the buffer molecule (see also \citep{Hatakeyama2012, Werner2015}).

From this scaling property of the buffered RD system, we can derive the reciprocal relationship between wavelength robustness and spatial-phase plasticity. When $w$ is small, the magnitude of the phase shift in space can be evaluated as below. When $\beta$ is transiently changed locally in space, the amplitude is altered locally. In this case, the relative change in the amplitude is proportional to $\Delta \ln w^* (\beta)$, since the concentration is scaled as $\tilde{x}_i = w^*(\beta) x_i$. Then, after a spatial perturbation, the spatial pattern relaxes to the steady state for the spatially uniform $\beta$. In a linear regime, the phase shift magnitude is proportional to the magnitude of the disturbance to the amplitude. Hence, phase changes are given as $\Delta \phi = a \Delta \ln w^* (\beta)$, where $a$ is a positive coefficient.

Therefore, changes in the wavelength and phase can be described as
\begin{equation}
\Delta \phi / a + 2 \Delta \ln \lambda = - \Delta \ln \mathscr{F} (\beta). \label{equalityspace}
\end{equation}
The right-hand side of Eq.(\ref{equalityspace}) depends only on $\Delta \ln \mathscr{F}(\beta)$, and not on $\Delta \ln w^*(\beta)$. From this equality, if we set the dynamics of buffer molecule to satisfy $\Delta \ln \mathscr{F}(\beta) = - \Delta \ln w^*(\beta)$, the wavelength is robust against environmental changes, i.e., $\Delta \lambda = 0$ and $\Delta \phi = - a \Delta \ln \mathscr{F} (\beta)$ is satisfied.
Thus, when the wavelength of a temporal pattern is robust, phase is plastic. Correspondingly, as the robustness of the wavelength is lost, the phase is less plastic. Eq.(\ref{equalityspace}) gives the reciprocity between wavelength robustness and spatial phase plasticity.

So far we did not include dependence of the diffusion constants upon the environmental parameter, which could be taken into account.
From the Stokes-Einstein law, it is natural to assume same dependence on the parameter (e.g., temperature) of the diffusion constants for different chemical species,
Then, the term $D_i \frac{\partial^2}{\partial {\bf r}^2} x_i$ in Eq.(\ref{dxdt}) will be replaced by $\mathscr{D}(\beta)  d_i \frac{\partial^2}{\partial {\bf r}^2} x_i$, where $\mathscr{D}(\beta)$ is the ESF for the diffusion constant.
In this case, by the similar scale transformation as above (${\bf r} \rightarrow {\boldsymbol \rho} \left\{\mathscr{D}(\beta) / \mathscr{F} (\beta) w^* (\beta) \right\}^{1/2}$, $t \rightarrow \tau / \mathscr{F} (\beta) w^* (\beta)$, $x_i \rightarrow \tilde{x}_i / w^* (\beta)$), the condition for robustness of the wavelength is obtained as $\Delta \ln \lambda = \left\{\Delta \ln \mathscr{D} (\beta) - \Delta \ln \mathscr{F} (\beta) - \Delta \ln w^* (\beta) \right\} / 2$, while the reciprocity between it and the phase plasticity is given by $\Delta \phi / a + 2 \Delta \ln \lambda = \Delta \ln \mathscr{D} (\beta) - \Delta \ln \mathscr{F} (\beta)$.

In general, each reaction term has different ESF in Eq.(\ref{dxdt}).
In this case, instead of considering the general form, consider 2-components RD system in the vicinity of a bifurcation of the Turing instability, where the characteristic wavelength is proportional to $(w^{*2} \mathrm{det} J)^{1/4}$, where $J$ is Jacobian of the original reactions depending on a few ESFs.
This form is derived from the linear stability analysis (for detailed calculation, see \citep{Supply}).
Then the condition for robustness is given by $2 \Delta \ln w^* + \Delta \ln \mathrm{det} J = 0$.
For example, if $f_1$ and $f_2$ have different ESFs, the product of two ESFs is factored out of the Jacobian, and if its dependence on $\beta$ is cancelled out by $w^*$, the condition is satisfied.

As an example of this general relationship, we study RD systems consisting of two components (an activator and an inhibitor) and one buffer. In particular, for numerical demonstration, we adopt the buffered Brusselator \citep{Prigogine1968} given as
\begin{subequations}
\begin{eqnarray}
\frac{\partial u}{\partial t} &=& \mathscr{F}(\beta) \left\{ A +  u_a^2  v_a - (B + 1) u_a \right\} + D_u \nabla^2 u,\\
\frac{\partial v}{\partial t} &=& \mathscr{F}(\beta) \left\{- u_a^2  v_a + B u_a \right\} + D_v \nabla^2 v, \\
\frac{\partial w}{\partial t} &=& g(w; \beta) + D_w \nabla^2 w,
\end{eqnarray}
\end{subequations}
where $u$, $v$, and $w$ are an activator, inhibitor, and buffer molecule, respectively.
We consider temperature change for our example of an environmental change and use the Arrhenius form for the ESF. We set $E_f$, $E_{g1}$, and $E_{g2}$ as activation energies for $\mathscr{F} (\beta)$, $\mathscr{G}_1 (\beta)$, and $\mathscr{G}_2 (\beta)$. We simulate wavelength robustness against the change in $\beta$ by using the 4th order Runge-Kutta method.

\begin{figure}[htbp]
\centering \includegraphics[width=8.7cm]{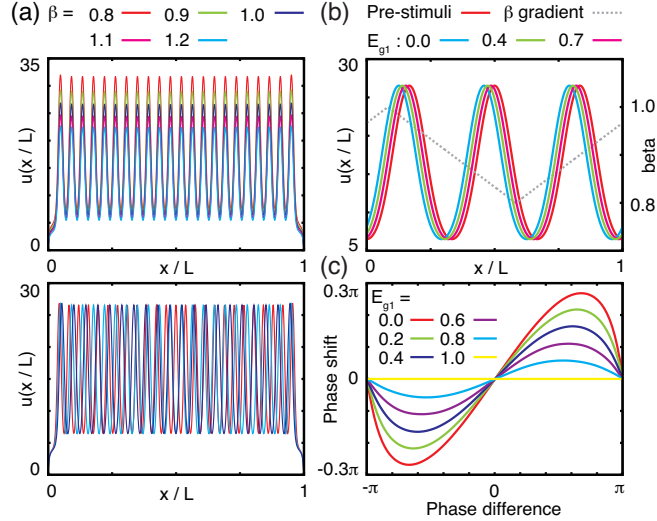}
\caption{
Wavelength robustness and spatial phase plasticity of the buffered Brusselator with $A=2$, $B=4$, $D_u=5$, and $D_v=30$.
(a) Environmental dependency of the wavelength under the Dirichlet boundary condition.
Different line colors indicate spatial patterns at different $\beta$, while $E_{g1}$ is set at 0 (up) and 1.0 (down).
Both of $E_{f}$ and $E_{g1}$ are fixed at 1.0.
ESFs are given as the Arrhenius function as $\mathscr{F}(\beta) = e^{E_f} \exp(-\beta E_f)$, $\mathscr{G}_1(\beta) = 3.0 e^{E_{g1}} \exp(-\beta E_{g1})$, and $\mathscr{G}_2(\beta) = e^{E_{g2}} \exp(-\beta E_{g2})$.
(b) Spatial phase shift after a stimulus, given as a transient change in $\beta$ from the spatially uniform $\beta = \beta_1$ to a gradually distributed $\beta$ between $\beta_1 = 1.0$ and $\beta_2 = 0.8$, shown as the gray dotted line, with duration time 100.0.
The red line is the spatial pattern before the stimulus, and the others indicate spatial patterns after the stimulus, for varied $E_{g1}$.
Periodic boundary condition is adopted to compute the phase shift of pattern.
(c) Spatial phase response curve (SPRC) of the buffered Brusselator against the stimulus.
The horizontal axis shows the phase differences between the spatial pattern generated by the buffered Brusselator and given by the $\beta$ gradient, as defined by the difference between the peaks of the pattern.
The applied $\beta$ gradient is normalized by the pattern wavelength.
Different colors indicate SPRCs for different $E_{g1}$ values.}
\label{robust_plast}
\end{figure}

If the concentration of buffer molecule at the steady state is independent of $\beta$, i.e., if $E_{g1}$ and $E_{g2}$ are same, the wavelength is not robust against changes in $\beta$. 
Instead, the amplitude of spatial pattern would be independent against changes in $\beta$.
Whereas, when $E_{g1}$ is small, the wavelength is robust across a wide range of $\beta$ and the amplitude is sensitive to changes in $\beta$ (see Fig.\ref{robust_plast}A).

\begin{figure}[htbp]
\centering \includegraphics{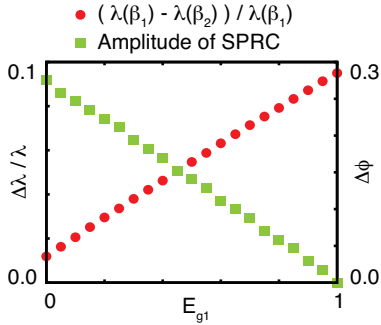}
\caption{
Reciprocity between wavelength robustness and spatial phase plasticity of the buffered Brusselator.
Red circles represent the difference in the wavelength at $\beta_1 = 1.0$ and at $\beta_2 = 0.8$ under the periodic boundary condition, normalized by the wavelength at $\beta_1$.
Green squares represent the difference between the maximal and minimal values of the SPRC, as shown in Fig.\ref{robust_plast}C.
}
\label{s_reciprocity}
\end{figure}

To study the effect of ESF on robustness quantitatively, we calculated the wavelength of systems with varied $E_{g1}$ at $\beta = \beta_1$ and $\beta = \beta_2$ ($\beta_1 > \beta_2$). We plot the difference between the wavelength at $\beta_1$ and $\beta_2$ ($\lambda_1$ and $\lambda_2$) as red circles in Fig.\ref{s_reciprocity}.
As $E_{g1}$ decreases, the wavelength difference monotonically decreases, i.e., the robustness of the wave number increases, while the sensitivity of the buffer molecule concentration increases, i.e., the amplitude is plastic against environmental changes.
Thus, changes in wavelength can be counterbalanced by changes in buffer molecule concentration.

Next, we studied the spatial phase shift against a transient change in environmental parameter $\beta$ as an indicator of plasticity. We transiently changed $\beta$ from the spatially uniform value $\beta_1$ to a non-uniform $\beta$ in space, taking a graded value between $\beta_1$ and $\beta_2$ (see the gray line in Fig.\ref{robust_plast}B). Then, $\beta$ is returned to $\beta_1$ uniformly in space, and the spatial pattern relaxes into the steady-state pattern with the original wavelength, while the phase of the pattern is shifted. The magnitude of the phase shift depends on the phase difference between the pattern generated by the RD system and the $\beta$ gradient. Thus, we plot a spatial phase response curve (SPRC) where the $x$-axis is the phase difference and the $y$-axis is the magnitude of the phase shift (Fig.\ref{robust_plast}C). The difference between the maximum and minimum values of the SPRC provides an indicator of phase plasticity.

If $E_{g1}$ is equal to $E_{g2}$, the buffer molecule concentration does not depend on $\beta$, and neither a phase shift nor wavelength robustness exists. Then, with the decrease in $E_{g1}$, the phase shift magnitude increases, as shown in Fig.\ref{s_reciprocity}, where the difference in phase is negatively correlated with the difference in the wavelength between two $\beta$ values. Moreover, the sum of the relative wavelength change $\Delta \lambda / \lambda (\beta_1)$ and the phase shift $\Delta \phi$ is almost constant within the whole range of $E_{g1}$ as $a \Delta \lambda / \lambda (\beta_1) + \Delta \phi = {\rm const.}$, where $a$ is a positive coefficient, as predicted.

RD systems often show temporal rhythms. Indeed the Belousov-Zhabotinsky reaction, which is the basis of the Brusselator, shows temporal oscillation in a well-mixed medium \citep{Belousov1959, Zhabotinsky1964} whereas it also can show a spatial pattern \citep{DeSimone1973}.
Moreover, some pattern formation mechanisms fixing the temporal rhythms are known, as in the clock and wavefront \citep{Cooke1976} and interaction-induced fixation of oscillation \citep{Cotterell2015, Kohsokabe2016}.
Previously we reported reciprocity between robustness of period and plasticity of phase in temporal oscillation, as mediated by the adaptation via buffer molecules.
Indeed, reciprocity for spatial pattern here and for the temporal rhythm are integrated below.

\begin{figure}[htbp]
\centering \includegraphics{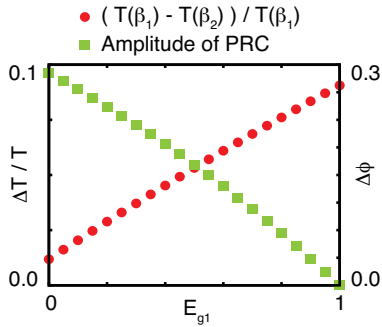}
\caption{
Reciprocity between the robustness of period and phase plasticity in the temporal rhythm of a buffered Brusselator. $A$ is set to 1.0 and other parameters are the same as the spatial pattern. $\Delta T / T$ is the difference of periods at $\beta_1 = 1.0$ and $\beta_2 = 0.8$ normalized by the period at $\beta_1$. The difference between the maximal and minimal values of the phase response curve is $\Delta \phi$ \citep{Winfree1980, Kuramoto1984}.
}
\label{t_reciprocity}
\end{figure}

From Eq.(\ref{scaledBuffer}), the timescale is proportional to $1 / \mathscr{F} (\beta) w^* (\beta)$ and the relative change in the period is given as $\Delta \ln T = - \left\{ \Delta \ln \mathscr{F} (\beta) + \Delta \ln w^* (\beta) \right\}$ owing to the rescaling of the timescale. Hence, the period is robust when $\Delta \ln \mathscr{F}(\beta) + \Delta \ln w^* (\beta) = 0$, which is same as the condition for wavelength robustness in the spatial pattern. Moreover, the phase shift is also proportional to a transient change in amplitude in the temporal rhythm, as given by $\Delta \phi = a' \Delta \ln w^* (\beta)$. Thus, reciprocity between periodic robustness and phase plasticity in time is given as
\begin{equation}
\Delta \phi / a' + \Delta \ln T = - \Delta \ln \mathscr{F} (\beta), \label{equalitytime}
\end{equation}
where $a'$ is a positive coefficient. Thus, reciprocity for the temporal rhythm is satisfied, because the robustness and reciprocity originate in the common scaling forms in Eq.(\ref{scaledBuffer}), for both time and space. For example, reciprocity in the temporal rhythm holds in the buffered Brusselator (see Fig.\ref{t_reciprocity}), as for spatial pattern. The scaling in the buffered RD system indicates robustness and reciprocity for any spatiotemporal pattern, including wave and scroll \citep{DeSimone1973, Siegert1995, Sawai2005}.

In this paper, we studied the relationship between wavelength robustness and phase plasticity in spatial pattern formation.
We demonstrated that the dynamics of a buffer molecule can counterbalance the environmental dependency of a spatial pattern generated by RD systems, and that wavelength robustness can be achieved via changes in the buffer molecule concentration.
Our mechanism works independently of the system size, while it is especially relevant for large systems against small environmental changes.
Simultaneously, the amplitude of spatial pattern changes following the change in buffer molecule concentration.
Thus, the phase of the spatial pattern is easily changed by a transient change in the environment.
Therefore, the reciprocity between wavelength robustness and spatial phase plasticity holds against the change in an environmental factor.
Previously, we demonstrated reciprocity between the periodic robustness and phase plasticity of circadian clocks \citep{Hatakeyama2015} in response to temperature change, due to the adaptation of a limit cycle by a buffering molecule \citep{Hatakeyama2012}, which could be interpreted as the scaling property, as discussed here. 
Here, it is interesting to note that adaptation dynamics sometimes follow a certain scaling property as, for example, discussed in fold-change detection \citep{Shoval2010}.
We expect that adaptation mechanisms due to the scaling property generally lead to reciprocity between robustness and plasticity.

In addition to robustness, phase plasticity in spatial pattern is important for biological organisms responding to environmental changes.
In taxis of cells, the angle of cellular polarity should be sensitive to changes in environmental factors \citep{Andrew2007}, whereas the number of the cellular compass should be robust.
Another example is in the differentiation of multicellular cyanobacteria, whereby the change in heterocyst position, and thus the efficiency of the division of labor between photosynthesis and nitrogen fixation are enhanced.
After transient starvation by nitrogen depletion, the excess heterocysts suppress growth. Hence, robustness in the fraction of heterocysts against nitrogen concentration is also important for effective growth in a fluctuating environment.
These examples suggest the relevance of reciprocity in spatial pattern to organismal fitness.
Further confirmation of the reciprocity between robustness and plasticity should be pursued, as well as confirmed experimentally.

\begin{acknowledgements}
This research is partially supported by the Platform for Dynamic Approaches to Living System from Japan Agency for Medical Research and Development from AMED, Japan, and JSPS KAKENHI Grant No. 15K18512 and 15H05746.
\end{acknowledgements}

\newpage

\begin{center}\
{\LARGE \textbf {Supplemental Material}}\\
{\large \textit{Robustness of spatial patterns in buffered reaction-diffusion systems and its reciprocity with phase plasticity}\\
Tetsuhiro S. Hatakeyama, Kunihiko Kaneko\\}
\end{center}

\newpage

Consider a reaction-diffusion system with two components in the vicinity of a bifurcation of the Turing instability, the linearized dynamics are given by
\begin{eqnarray*}
\frac{\partial x_1}{\partial t} = f_{1,1} (\beta) x_1 + f_{1,2} (\beta) x_2 + D_{x_1}\frac{\partial^2 x_1}{\partial {\bf r}^2}, \\
\frac{\partial x_2}{\partial t} = f_{2,1} (\beta) x_1 + f_{2,2} (\beta) x_2 + D_{x_2} \frac{\partial^2 x_2}{\partial {\bf r}^2},
\end{eqnarray*}
where $f_{1,1} (\beta) = \left. \frac{\partial f_1 (x_1,x_2)}{\partial x_1} \right|_{\{x_1, x_2\} = \{x_1^*, x_2^*\}}$, $f_{1,2} (\beta) = \left. \frac{\partial f_1 (x_1,x_2)}{\partial x_2} \right|_{\{x_1, x_2\} = \{x_1^*, x_2^*\}}$, $f_{2,1} (\beta) = \left. \frac{\partial f_2 (x_1,x_2)}{\partial x_1} \right|_{\{x_1, x_2\} = \{x_1^*, x_2^*\}}$, $f_{2,2} (\beta) = \left. \frac{\partial f_2 (x_1,x_2)}{\partial x_2} \right|_{\{x_1, x_2\} = \{x_1^*, x_2^*\}}$.
In this case, the characteristic wavenumber is given as
\begin{equation*}
k = \left( \frac{f_{1,1} f_{2,2} - f_{1,2} f_{2,1}}{D_{x_1} D_{x_2}} \right)^{1/4}.
\end{equation*}

When only the active molecule, which is given by $x_1^a = w^* x_1$ and  $x_2^a = w^* x_2$ where $K_{x_1}$ and $K_{x_2}$ are set as 1 for convenience, can be incorporated into reactions, the linearized dynamics are given by
\begin{eqnarray*}
\frac{\partial x_1}{\partial t} &=& \left. \frac{\partial f_1}{\partial x_1^a} \right|_{\{x_1^a, x_2^a\} = \{x_1^{a*}, x_2^{a*}\}} \frac{\partial x_1^a}{\partial x_1}  x_1 + \left. \frac{\partial f_1}{\partial x_2^a} \right|_{\{x_1^a, x_2^a\} = \{x_1^{a*}, x_2^{a*}\}} \frac{\partial x_2^a}{\partial x_2}  x_2  + D_{x_1} \frac{\partial^2 x_1}{\partial {\bf r}^2}  \\
&=&  f_{1,1} (\beta) w^* (\beta) x_1 + f_{1,2} (\beta) w^* (\beta) x_2 + D_{x_1} \frac{\partial^2 x_1}{\partial {\bf r}^2}, \\
\frac{\partial x_2}{\partial t} &=&  \left. \frac{\partial f_2}{\partial x_1^a} \right|_{\{x_1^a, x_2^a\} = \{x_1^{a*}, x_2^{a*}\}} \frac{\partial x_1^a}{\partial x_1} x_1 + \left. \frac{\partial f_2}{\partial x_2^a} \right|_{\{x_1^a, x_2^a\} = \{x_1^{a*}, x_2^{a*}\}} \frac{\partial x_2^a}{\partial x_2} x_2  + D_{x_2} \frac{\partial^2 x_2}{\partial {\bf r}^2} \\
&=&  f_{2,1} (\beta) w^* (\beta) x_1 + f_{2,2} (\beta) w^* (\beta) x_2 + D_{x_2} \frac{\partial^2 x_2}{\partial {\bf r}^2}.
\end{eqnarray*}
Hence, the characteristic wavenumber is given as
\begin{equation*}
k = \left\{ \frac{w^{*2} (f_{1,1} f_{2,2} - f_{1,2} f_{2,1})}{D_{x_1} D_{x_2}} \right\}^{1/4}.
\end{equation*}
Therefore, the condition for robustness by scaling is relaxed into
\begin{equation*}
2 \Delta \ln w^* (\beta) + \Delta \ln (f_{1,1} (\beta) f_{2,2} (\beta) - f_{1,2} (\beta) f_{2,1} (\beta)) = 0,
\end{equation*}
and then the all of reactions do not have to show the same environmental dependency.
For example, if $f_1$ and $f_2$ have different ESFs, the product of two ESFs is factored out of the Jacobian, and if its dependence on $\beta$ is cancelled out by $w^*$, the condition is satisfied.
As another example, when activation and inhibition reactions are respectively catalyzed by different enzymes and each of the enzymes has different EFS, i.e., EFS for $f_{1,1}$ and $f_{2,1}$ are common, that of $f_{1,2}$ and $f_{2,2}$ are also common (but can be different from the former), the product of two ESFs is also factored out and the above condition will be satisfied.


\begin{thebibliography}{99}
\bibitem{foot1}
This transform might change a bifurcation point against changes in $K_{x_i}$s, but our discussion is not affected by it.
\end{thebibliography}
\end{document}